\documentclass[aps,prb,twocolumn,showpacs,superscriptaddress]{revtex4}

\usepackage{graphicx} 

\begin{document}

\title
{Adsorption and dissociation of hydrogen molecules on bare and
functionalized  carbon nanotubes}

\author{S. Dag}
\affiliation{Department of Physics, Bilkent University, Ankara
06800, Turkey}
\author{Y. Ozturk}
\affiliation{Department of Physics, Bilkent University, Ankara
06800, Turkey}
\author{S. Ciraci}
\affiliation{Department of Physics, Bilkent University, Ankara
06800, Turkey}
\author{T. Yildirim}
\affiliation{NIST Center for Neutron Research, Gaithersburg, MD 20899}

\date{\today}

\begin{abstract}
We investigated interaction between hydrogen molecules and bare as
well as  functionalized single-wall carbon nanotubes (SWNT) using
first-principles plane wave method. We found that the binding
energy of the H$_{2}$ physisorbed on the bare SWNT is very  weak,
and can be enhanced neither by increasing the curvature of the
surface through radial deformation, nor by the coadsorption of Li
atom that makes the semiconducting tube metallic. Though the
bonding is strengthened upon adsorption directly to Li atom, yet
its nature continues to be physisorption. However, the character
of the bonding changes dramatically when SWNT is functionalized by
the adsorption of Pt atom. Single H$_{2}$ is chemisorbed  to Pt
atom on the SWNT either dissociatively or molecularly. If Pt-SWNT
bond is weakened either  by displacing Pt from bridge site to a
specific position or by increasing number of the adsorbed H$_{2}$,
the dissociative adsorption of H$_{2}$ is favored. For example,
out of two adsorbed H$_{2}$, first one can be adsorbed
dissociatively, second one is chemisorbed molecularly. The nature
of bonding  is weak physisorption for the third adsorbed H$_{2}$.
Palladium also promotes the chemisorption of H$_{2}$ with
relatively smaller binding energy. Present results reveal the
important effect of transition metal atom adsorbed on SWNT and
advance our understanding of the molecular and dissociative
adsorption of hydrogen  for efficient hydrogen storage.

\end{abstract}

\pacs{73.22.-f, 61.46.+w, 68.43.Bc}

\maketitle

\section{Introduction}
 Fuel cells have been a real challenge for clean and efficient
source of energy in diverse fields of applications with different
size and capacity range. Once hydrogen molecule is chosen as
potential fuel, its storage, easy discharge for consumption and
dissociation into hydrogen atoms to complete the current cycle in
the fuel cell to produce the desired electromotive force involve
several problems to be solved. Dillon {\it et al.}\cite{dillon}
have pioneered the idea that carbon nanotubes can be efficient,
cheap and rechargeable storage medium for small-scale fuel cells
by estimating 5-10 weight percent (\emph{wp}) H$_{2}$ adsorption
in single-wall carbon nanotubes (SWNT). Later, Ye \textit{et
al.}\cite{ye} and Liu \textit{et al.}\cite{liu} obtained H$_{2}$
storage capacities of 8.2 and 4.2 \emph{wp}, respectively.
Unfortunately, recent studies further exploring this idea have
come up with controversial
conclusions\cite{lee,ma,darkrim,eklund,shiraishi,chen}. In the
mean time, adsorption of alkali atoms  on SWNTs have been proposed
to enhance the H$_{2}$-uptake\cite{chen,dubot}. Nevertheless,
carbon nanotubes have high surface-volume ratio and their
functionalization to render them feasible for hydrogen storage
through coverage of suitable adatoms has remained to be explored.

In this paper\cite{yavuz}, we addressed  following questions in
order to clarify controversial issues related to the storage of
hydrogen molecule on carbon nanotubes: Can H$_{2}$ molecule be
adsorbed on the SWNT? What is the nature and strength of the
bonding? Can the strength of the bonding be modified either by
changing the curvature of the surface or by the coadsorption of
metal atoms? In particular, can the H$_{2}$-uptake on SWNTs be
promoted by their functionalization through transition elements?
How can H$_{2}$ molecule be dissociated? To answer all these
questions we investigated the interaction between H$_{2}$ molecule
and bare, radially deformed and foreign atom adsorbed SWNTs by
carrying out calculations within Density Functional Theory
(DFT)\cite{dft}. Our results not only  advance our understanding
of H$_{2}$ adsorption on  carbon nanotubes, but also suggest new
ways for efficient hydrogen storage for rapidly growing research
on fuel cells.

\begin{figure}
\includegraphics[scale=0.5]{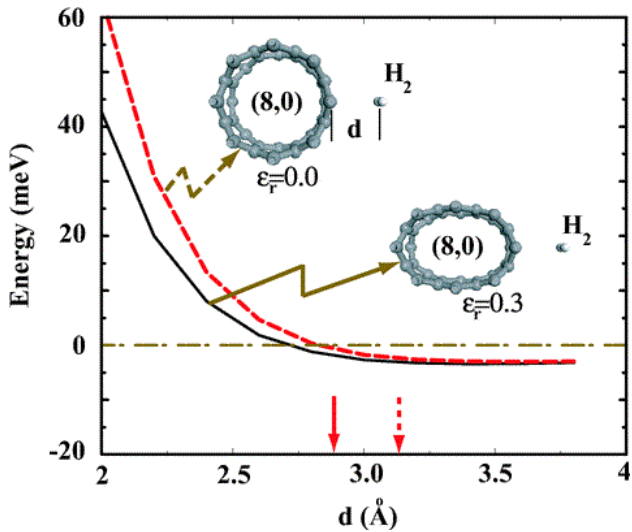}
\caption{Variation of chemical interaction energy $E_{C}$ between
SWNT and H$_{2}$ molecule as a function of distance $d$ between
them. Two cases, namely adsorption to bare and radially deformed
SWNT are shown by dashed and continuous lines, respectively. In
calculating both curves, atomic structures corresponding to $d
\rightarrow \infty$ have been used without relaxation. Dash-dotted
line indicate zero of chemical interaction energy. Optimized
distance for two cases are indicated by arrows.} \label{fig:bare}
\end{figure}

Our calculations have been carried out using first-principles
plane wave method and ultrasoft
pseudopotentials\cite{vander,hafner} within Generalized Gradient
Approximation (GGA)\cite{perdew}. Adsorption and dissociation of
H$_{2}$ is treated within the supercell geometry by optimizing
atomic positions. The weak attractive Van der Waals (VdW)
interaction becomes crucial in  calculating binding energies of
weak physisorption bonds, but is not well represented in DFT using
GGA\cite{kohn}. Therefore, in the case of physisorption, weak and
attractive VdW interaction energy, $E_{VdW}$, is obtained from the
Slater-Kirkwood approximation\cite{slater} using the asymptotic
form of the Lifshitz's equation\cite{lifshitz}. However, this
approach can not be suitable to determine the contribution of VdW
interaction in the chemisorption of molecules. In the present
calculations we take the zigzag (8,0) SWNT as a prototype tube.

\section{ Adsorption of H2 on bare and radially deformed SWNT}
 To clarify whether H$_{2}$ can form stable bonding on the
outer or inner surface of a SWNT, we calculated the chemical
interaction energy between H$_{2}$ and the outer surface of the
(8,0) SWNT at different sites (\textit{i.e.} H-site, above the
hexagon; Z- and A-site above the zigzag and above the axial C-C
bonds; T-site, a bridge site between two adjacent zigzag C-C
bonds) as a function of spacing $d$. The chemical interaction
energy is obtained from the expression,
$E_{C}(d)=E_{T}[H_{2}+SWNT,d]-E_{T}[SWNT]-E_{T}[H_{2}]$, in terms
of the total energies of  bare nanotube ($E_{T}[SWNT]$), free
H$_{2}$ ($E_{T}[H_{2}]$), and H$_{2}$ attached to SWNT at a
distance $d$ ($E_{T}[H_{2}+SWNT,d]$). Here $E_{C}<0$ corresponds
to an attractive interaction. The stable binding occurs at the
minimum of $E_{C}(d)+E_{VdW}(d)$, the negative of it is denoted as
the binding energy $E_{b}$. The binding is exothermic when
$E_{b}>0$. In Fig.~\ref{fig:bare} we show the variation of
$E_{C}(d)$ calculated for unrelaxed atomic structures at the
H-site. Once the atomic structure of both SWNT and H$_{2}$
molecules are relaxed the mimimum value of $E_{C}(d)$ is found to
be -27 meV at $d_{0}=3.1$\AA{} at the H-site. Minimum values of
$E_{C}(d)$ calculated for A-, Z-, and T-sites are also very small
and comparable to that of H-site. The long range VdW interaction
energy calculated for the H-site at $d_{o}$ is $E_{VdW}\sim $-30
meV. Then the binding energy associated with H$_{2}$ molecule
adsorbed at H-site is calculated to be $E_{b}\sim $57 meV. This is
a small binding energy and indicates physisorption.

\begin{figure}
\includegraphics[scale=0.5]{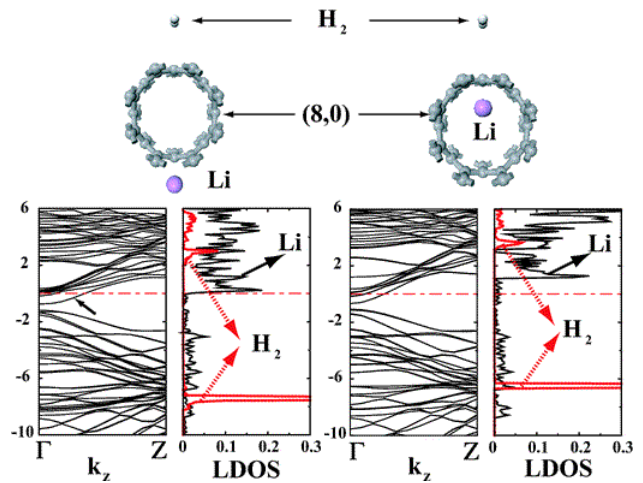}
\caption{ Atomic configuration, energy band structure and local
density of states (LDOS) calculated for the coadsorption of
H$_{2}$ molecule and single Li atom. Two cases correspond to Li
atom chemisorbed on the external and internal surface of the (8,0)
zigzag SWNT. Zero of energy is set at the Fermi level, $E_{F}$.
LDOS calculated at Li and H$_{2}$ are shown by continuous and
dotted lines, respectively. Metallized SWNT bands are indicated by
arrows.} \label{fig:lih}
\end{figure}

Previous studies\cite{lee,ma} revealed that the physisorption of
individual H$_{2}$ molecules with $E_{b}>0$ cannot occur on the
inner wall of SWNT. Hydrogen molecules prefer to stay either at
the center of the tube or to  form some cylindrical shells inside
depending on radius of the tube. Owing to increased H-H
interaction at high coverage, even the atomic hydrogen cannot form
stable structure when it is adsorbed to the inner  wall of small
radius tube. For (8,0) tube we found that H$_{2}$ is trapped and
stabilized at the center of the tube with a repulsive interaction
energy $E_{C}=+0.34$ eV. The implementation of H$_{2}$ inside the
tube having radius in the range of 3 \AA~ is expected to be
hindered by this repulsive interaction.

Earlier it has been shown that binding energy of foreign atoms
adsorbed on SWNT increase with increasing curvature\cite{oguz}.
Tada \textit{et al.}\cite{tada} have argued that the potential
barrier associated with the dissociative adsorption of H$_{2}$ on
SWNT is lowered with increasing curvature of the tube. It has been
proposed that the potential barrier for the dissociation of
H$_{2}$ adsorbed in the interstitial region between tubes can be
lowered by applying radial deformation to the rope or
SWNT\cite{chan}. It is known that under radial deformation the
circular cross section changes and consequently the curvature
varies at different locations on the surface. Motivated with those
effects of curvature, we examine whether the attractive
interaction energy $E_{C}$ can be enhanced by changing the
curvature of the tube via radial deformation. Radial deformation
is realized by pressing the tube between two ends of a given
diameter. It, in turn, changes the circular cross section of the
bare tube with radius $R_{o}$ to an elliptical one with major and
minor axes $2a$ and $2b$, respectively. The atomic structure of
(8,0) tube is optimized under radial strain
$\epsilon_{r}=(R_{o}-b)/R_{o}\simeq $0.3 by fixing row of carbon
atoms at the end of minor axis. The deformation is reversible so
that the tube goes back to its original, undeformed form upon the
release of radial strain\cite{oguz2}. The deformation energy (that
is the difference between the total energies of deformed and
undeformed SWNTs) is calculated to be $E_{D}=$1.4 eV per unit
cell. We examined whether  the binding energy of H$_{2}$ molecule
changes under the radial deformation of SWNT.
Figure~\ref{fig:bare} shows the variation of $E_{C}(d)$ for
H$_{2}$ approaching toward the high curvature site of the tube
(\textit{i.e.} one end of the major axis) at the H-site. The
minimum value of the attractive interaction,  $E_{C}$ is -30 meV
and  occurs at $d_{o}$=2.9 \AA. For $\epsilon_{r}=$0.25, we also
obtained very small enhancement of $E_{C}(d)$. Hence, the effect
of curvature on the binding energy of H$_{2}$ is negligible due to
relatively large $d_{o}$. This result also suggests that the
physisorption energy does not vary significantly depending on the
radius of SWNT. Apparently, the binding of H$_{2}$ on the outer
surface of SWNT is weak and corresponding physisorption energy is
small. The binding cannot be enhanced by increasing the curvature
locally through radial deformation. Curvature effect or radial
deformation may be significant at small $d$ when H$_{2}$ is forced
towards SWNT surface.

\section{Coadsorption of hydrogen molecule and Lithium atom on SWNT}
 Next, we examine whether the binding of H$_{2}$ is enhanced by the
coadsorbed foreign atoms. To this end we first consider  Li atom
adsorbed  on the (8,0) SWNT, since  the adsorption of an alkali
atom has been proposed to enhance the
H$_{2}$-uptake\cite{chen,dubot}. Li atom is chemisorbed at the
H-site, 1.5 \AA~ above the surface of SWNT with a binding energy
of 0.8 eV. Self-consistently calculated electronic structure shown
in Fig.~\ref{fig:lih} reveals that chemisorbed Li atoms donate
their 2$s$-valence electrons to the lowest conduction
$\pi^{*}$-band so that the semiconducting (8,0) SWNT (having band
gap $E_{g}=$0.6 eV) becomes metallic. This is a behavior common to
the other alkali atoms adsorbed on SWNTs\cite{engin}. In order to
examine the indirect effect of coadsorbed Li we considered H$_{2}$
as attached at the opposite site to Li. The optimized structure of
the physisorbed H$_{2}$ is shown in Fig.~\ref{fig:lih} together
with relevant structural parameters. We found $E_{C}$ has a
minimum value of -35 meV at $d_{o}=$3.4 \AA. Similar study has
been also performed for Li atom adsorbed on the inner wall of SWNT
while H$_{2}$ is on the external wall directly above the
coadsorbed Li as shown in Fig.~\ref{fig:lih}. In this adsorption
configuration minimum value of $E_{C}$ practically did  not
change. The local density of states calculated on Li atom and
H$_{2}$ refuse the possibility of any significant interaction
between adsorbates. As a result, our calculations for both
external and internal adsorption of Li rule out any indirect
effect of coadsorbed Li to enhance the binding of H$_{2}$ on SWNT.
The occupation of empty conduction band by the alkali electrons
and hence metallization of SWNT did not affect the bonding of
H$_{2}$. These results are in agreement with the first principles
calculations by Lee \textit{et al.}\cite{eclee}. However, the
effect of Li on the adsorption of H$_{2}$, whereby H$_{2}$ is
attached  directly to  Li atom is found significant. The minimum
value of $E_{C}$ has increased to -175 meV, while $d_{o}$
decreased to 2.1 \AA. Briefly, the coadsorption of Li does not
have any indirect effect on the binding of H$_{2}$, but the energy
associated with direct binding to Li is enhanced. However, the
nature of bonding remains physisorption in direct and indirect
cases.

\begin{figure}
\includegraphics[scale=0.55]{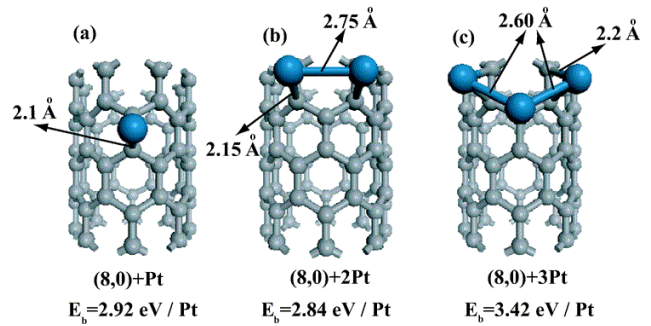}
\caption{(a) Atomic configurations for single, double and triple
Pt atoms adsorbed on the (8,0) SWNT. Average binding energy of
adsorbed Pt atoms $E_{b}$ and bond distances are indicated.}
\label{fig:ptconf}
\end{figure}

\begin{figure}
\includegraphics[scale=0.55]{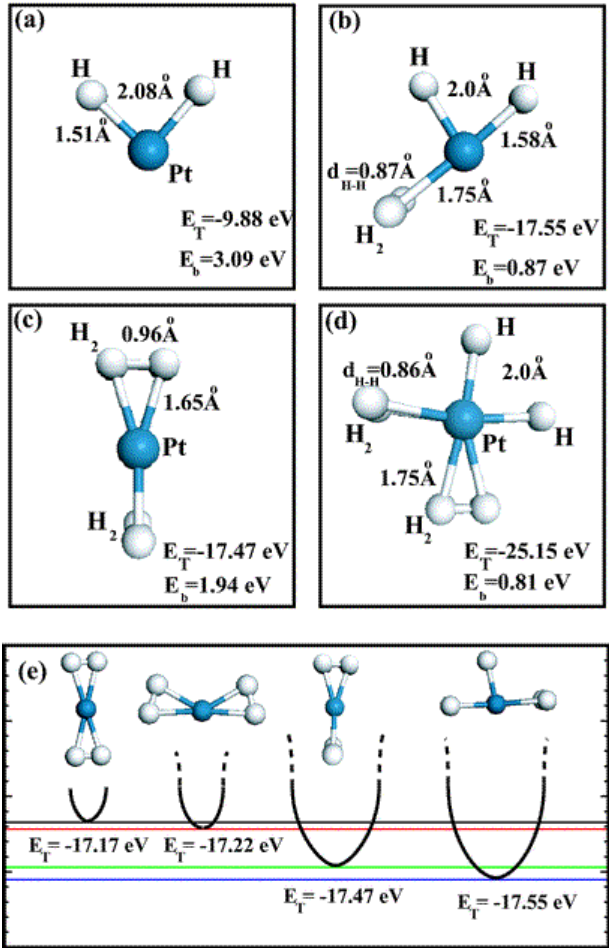}
\caption{Optimized binding configuration of H$_{2}$ molecules
adsorbed to a free Pt atom. (a) Dissociative adsorption of a
single H$_{2}$ molecule. (b) The first H$_{2}$ is dissociatively,
second H$_{2}$ molecularly adsorbed. (c) Two H$_{2}$ are
molecularly adsorbed. (d) Two H$_{2}$ are molecularly, one H$_{2}$
dissociatively adsorbed (e) Four different configurations related
with the adsorption of two H$_{2}$ to the same free Pt atom and
corresponding  four minima on the Born-Oppenheimer surface
described schematically. Binding energy of each adsorbed
additional hydrogen molecule E$_{b}$, total energy with respect to
constituent atoms E$_{T}$ and bond distances are indicated. }
\label{fig:freept}
\end{figure}

\section{Coadsorption of Hydrogen molecule and Platinum atom on SWNT}

A single transition metal atom adsorbed on the outer surface of
SWNT has shown interesting properties, such as high binding energy
and magnetic ground state with high net magnetic moment. For
example, transition element atoms (Ti, V, Cr, Mn, Fe, Co, Pt,
\emph{etc.}) have crucial adsorption states on
nanotubes\cite{engin} and some of them (Ti, Ni, Pd) form
continuous or quasi-continuous metal coating on the
SWNT\cite{zhang,sefa2}. As for Pt atom, it is known to be a good
cathalist in various chemical processes. While SWNTs offer high
surface/volume ratio, the interaction between H$_{2}$ and Pt atom
adsorbed on SWNT may be of interest. Now we investigate the
character of the bonding between H$_{2}$ and Pt adsorbed on SWNT
and address  the question of how many H$_{2}$ molecules can be
attached to an adsorbed Pt atom and how strong is the binding.

\subsection{Adsorption of  Pt atoms on SWNT}

 We first examine the
adsorption of Pt atom(s) on (8,0) SWNT. The character of the
bonding  has been investigated by placing Pt atoms on the A-sites
 of (8,0) tube (that is known to yield highest binding energy\cite{engin}) and
then by optimizing the structure. Three different adsorption
configuration have been examined, namely one, two and three Pt
atoms adsorbed on the adjacent sites to represent a small cluster
on SWNT as described in Fig.~\ref{fig:ptconf}. Calculated binding
energies of Pt atoms have increased as the number of Pt atoms
increases from one to three in the same neighborhood.  On the
other hand, the C-Pt distance gradually increases with increasing
number of Pt atoms adsorbed in the same neighborhood. This
paradoxical situation can be understood by the increasing Pt-Pt
coupling, which happens to be comparable to  C-Pt coupling derived
by the Pt-3$d$ and C-2$p$ orbitals.

\begin{figure}
\includegraphics[scale=0.55]{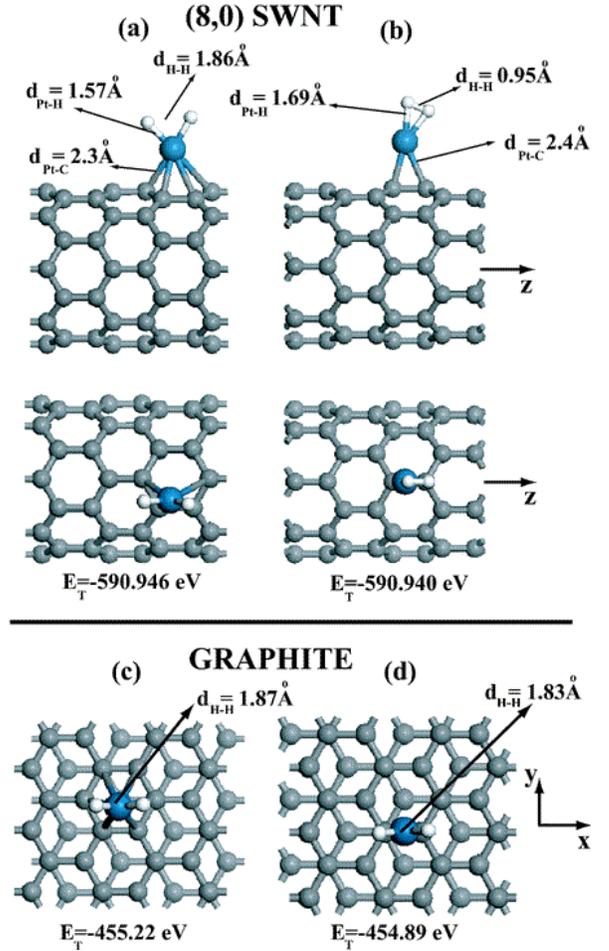}
\caption{Optimized adsorption geometry of H$_{2}$ molecule  on  a
single Pt atom. (a) Pt atom is adsorbed near the $H$-site of (8,0)
SWNT (side and top view) (b) Pt at the $A$-site (bridge position)
of (8,0) SWNT (side and top view) (c) Pt atom is adsorbed near the
$H$-site of the graphite surface (d) Pt atom at the $A$-site of
graphite. E$_{T}$ is the total energy relative to the constituent
C, Pt and H atoms.} \label{fig:pth}
\end{figure}

\subsection{Adsorption of H2 to a free Pt atom}
We now  consider   the interaction between H$_{2}$ molecules and a
free Pt atom. Optimized binding structures are shown in
Fig.~\ref{fig:freept}. Upon approaching to a free Pt atom, single
H$_{2}$ molecule dissociates and form PtH$_{2}$ with Pt-H bond
distance $d_{Pt-H}$=1.51 \AA{} and $d_{H-H}$=2.08 \AA. The total
energy E$_{T}$ relative to the energies of free Pt and H atoms is
calculated -9.88 eV. The binding energy relative to H$_{2}$
molecule and free Pt atom is E$_{b}$=3.09 eV. As for the
adsorption of H$_{2}$ to PtH$_{2}$, there are several minima on
the Born-Oppenheimer surface: The first minimum corresponds to a
configuration in Fig.~\ref{fig:freept}(b) where PtH$_{2}$
preserves the dissociated configuration while second H$_{2}$ is
molecularly adsorbed. Even if H-H interaction of adsorbed H$_{2}$
is weakened and hence the H-H distance has increased to 0.87\AA,
we identify it as molecular adsorption.  We denote this
configuration as PtH$_{2}$+H$_{2}$. The adsorbed H$_{2}$ molecule
is perpendicular to the plane of PtH$_{2}$. The binding energy of
the second H$_{2}$ to PtH$_{2}$ is calculated to be E$_{b}$=0.87
eV. Under these circumstances the average binding energy of each
H$_{2}$ is 1.98 eV. In the second configuration, identified as
Pt-2H$_{2}$ as shown in Fig.~\ref{fig:freept}(c), both H$_{2}$ are
molecularly adsorbed. As H-H molecular bonds are
 weakened,  $d_{H-H}$ is increased to 0.96\AA{} and all Pt-H
bonds have uniform  length with $d_{Pt-H}$ =1.65\AA. Here adsorbed
H$_{2}$ molecules are perpendicular. The binding energy of each
molecules is calculated to be 1.94 eV slightly less then the
average binding energy in PtH+H$_{2}$ configuration. The
configuration PtH$_{2}$+2H$_{2}$ shown in Fig.~\ref{fig:freept}(d)
involves the adsorption of three H$_{2}$ molecules; one
dissociatively, remaining two are molecularly adsorbed. Here
Pt-H$_{2}$ planes of two molecularly adsorbed H$_{2}$ are
perpendicular. The binding energy of the third H$_{2}$ relative to
the energy of PtH$_{2}$+H$_{2}$ in Fig.~\ref{fig:freept}(b) is
found to be 0.81 eV. Accordingly, the average binding energy of
each H$_{2}$ is 1.6 eV relative to free H$_{2}$ and free Pt atom.
Fig.~\ref{fig:freept}(e) compares four distinct configurations
related with the adsorption of two molecules on the same free Pt
atom. It appears that these configurations correspond to local
minima on the Born-Oppenheimer surface and the configuration in
Fig.~\ref{fig:freept}(b) appears to have lowest energy.

\begin{figure}
\includegraphics[scale=0.75]{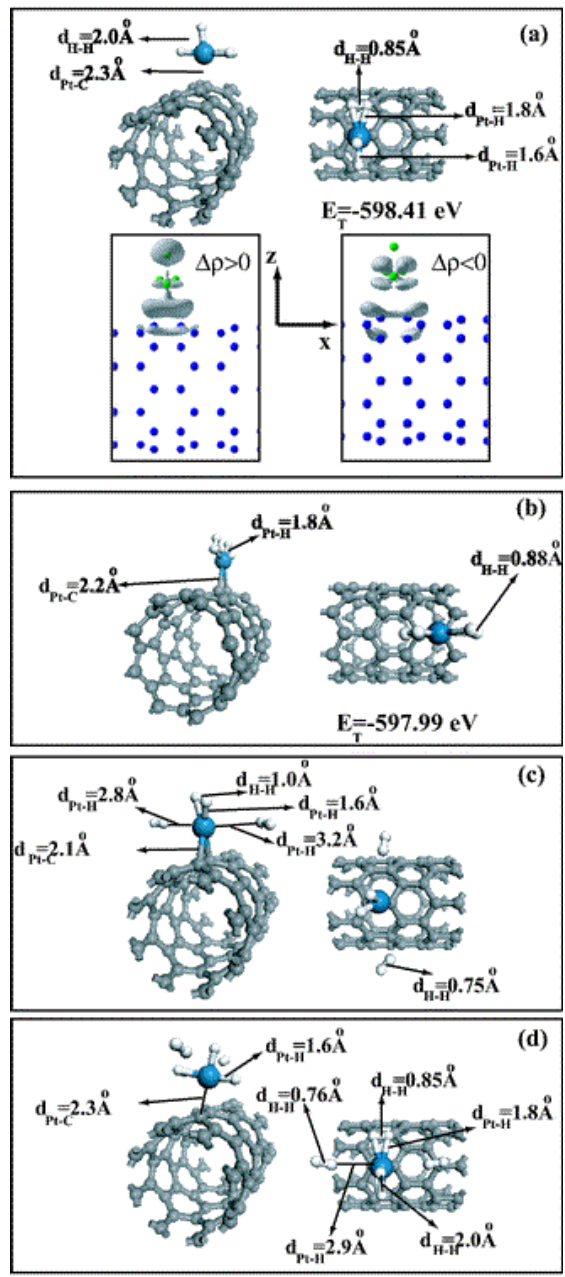}
\caption{ Optimized structure of H$_{2}$ molecules adsorbed to the
Pt atom on the SWNT. (a) One H$_{2}$ adsorbed to PtH$_{2}$. The
inset show the regions of charge depletion ($\Delta \rho <0$) and
charge accumulation ($\Delta \rho
> 0$) as a result of the bonding between SWNT and
PtH$_{2}$+H$_{2}$ in (a). (b) Another local minima where two
H$_{2}$ is molecularly adsorbed to the Pt atom. (c) One H$_{2}$ is
chemisorbed two H$_{2}$ are weakly bound. (d) four H$_{2}$. }
\label{fig:pt1h2}
\end{figure}

\subsection{Adsorption of H2 to a Pt atom on SWNT}

We deduced two configurations for the adsorption of a single
molecule to a Pt atom adsorbed on the (8,0) SWNT as described in
Fig.~\ref{fig:pth}. While these two chemisorption configuration
look dramatically different, their total energies differ only by 6
meV (that is smaller than the accuracy range of DFT). In the
configuration described in Fig.~\ref{fig:pth}(a) H$_{2}$ is
dissociatively adsorbed with binding energy E$_{b}$=1.1 eV
relative to free H$_{2}$ and Pt adsorbed on SWNT, \textit{i.e.}
E$_{T}$[H$_{2}$] and E$_{T}$[Pt+SWNT], respectively. The H-H and
Pt-H distances are 1.86 \AA{} and 2.3 \AA{}, respectively. Whereas
in the configuration  in Fig.~\ref{fig:pth}(b) H$_{2}$ is
molecularly adsorbed with a significantly  weakened H-H bond.
H$_{2}$ approaching from different directions and angles results
in a chemisorption state with binding energy $E_{b}=$1.1 eV and
Pt-H distance 1.7 \AA. The length of H-H bond has increased from
0.75 \AA{} to 0.95 \AA{} upon adsorption\cite{hydrogen}. Whereas,
this configuration appears to be less energetic than the former
one, the energy difference between them is only 6 meV. Notably,
while in the first configuration leading to dissociative
adsorption Pt atom is located near hollow H-site, in the
"molecular" adsorption of H$_{2}$ Pt atom is adsorbed at the
A-site. As compared to the average binding energy calculated for
H$_{2}$ adsorption to a free Pt atom,  the molecular adsorption of
H$_{2}$ to Pt atom adsorbed on SWNT is relatively weaker due to
the Pt-SWNT bond.

Adsorption of H$_{2}$ to a single Pt atom attached to the surface
of graphite is of interest in order to reveal how the binding
energy and binding configuration of H$_{2}$ depends on the radius
of SWNT. We considered two configuration, namely single Pt atom is
adsorbed near hollow H-site, as shown in Fig.~\ref{fig:pth}(c) and
Pt at the A-site as shown in Fig.~\ref{fig:pth}(d). For both
location of Pt atom on the graphite surface,  H$_{2}$ molecule
approaching the adsorbed Pt atom is dissociated and eventually Pt
atom formed two Pt-H bonds with individual H atoms. In this case
the binding with the graphite surface is weaker than that on SWNT,
and thus $d_{Pt-C}$ is increased to 2.4\AA. Relatively weaker
interaction between Pt and graphite surface allows  stronger
interaction between H$_{2}$ and Pt, as in the case of free Pt
atom, and hence leads to  the dissociation of the molecule. In
view of two limiting case in Fig.~\ref{fig:pth}, one can expect
that dissociation of H$_{2}$ may occur for SWNTs having larger R.

\begin{figure*}
\includegraphics[scale=0.88]{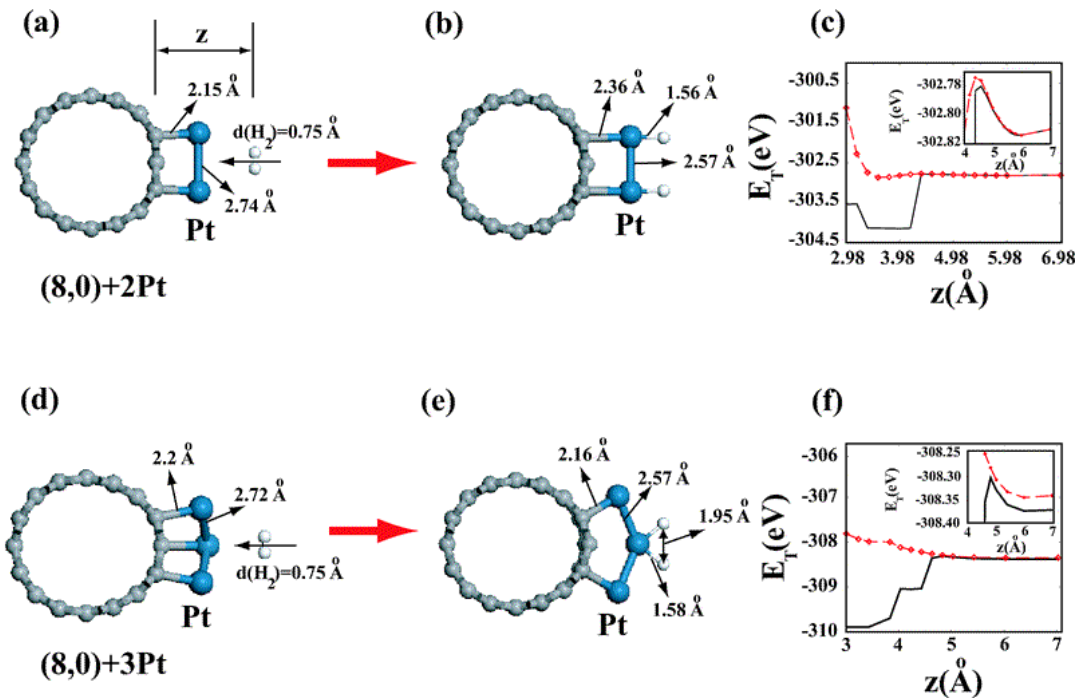}
\caption{Dissociative adsorption of single H$_{2}$ on a small Pt
cluster adsorbed on SWNT (a). One H$_{2}$ is approaching two
adjacent Pt atoms adsorbed on SWNT. (b) Optimized geometry after
dissociative adsorption of H$_{2}$. (c) Variation of total energy
with distance $z$. Small circles corresponds to E$_{T}$ for
unrelaxed H$_{2}$ and unrelaxed SWNT. Dotted curve corresponds to
E$_{T}$ of  the geometry relaxed  at certain distances. (d), (e)
and (f) same as (a), (b) and (c) except the Pt cluster consists of
3 Pt atoms. $z$ is the distance from the surface of SWNT.
Variation of E$_{T}(z)$ is amplified by inset. } \label{fig:ptall}
\end{figure*}

To answer the questions of  how many H$_{2}$ molecule can be
attached to a single Pt atom we perform a systematic study
outlined in Fig.~\ref{fig:pt1h2}. First we let a second H$_{2}$
approaches to Pt atom on SWNT that already has one  H$_{2}$
molecule attached as in Fig.~\ref{fig:pth}(b). The final optimized
geometry of Pt and adsorbed H$_{2}$ molecules in
Fig.~\ref{fig:pt1h2}(a) is similar to the configuration
PtH$_{2}$+H$_{2}$ described in Fig.~\ref{fig:freept}(b). This
situation is explained by the weakening Pt-C bond between
PtH$_{2}$ and SWNT shown in Fig.~\ref{fig:pth}(a)  which is
followed by the increase of $d_{Pt-C}$. First H$_{2}$, which was
initially chemisorbed to Pt as a molecule has dissociated upon the
molecular adsorption of the second H$_{2}$. The dissociation of
H$_{2}$ is an indirect process and is mediated by the weakening of
the Pt-C bonding through the molecular adsorption of second
H$_{2}$. The difference charge density
$\Delta\rho(\rm{\textbf{r}})=\rho_{T}(\rm{\textbf{r}})-
\rho_{SWNT}(\rm{\textbf{r}})-\rho_{PtH_{2}+H_{2}}(\rm{\textbf{r}})$
 calculated from the difference of total charge density
$\rho_{T}(\rm{\textbf{r}})$ of SWNT+PtH$_{2}$+H$_{2}$ in
Fig.~\ref{fig:pt1h2}(a) and those
 of SWNT and PtH$_{2}$+H$_{2}$ indicates that while the charge of
 Pt-$d_{xy}$ and C-$p_{x,y}$ orbitals are depopulated, the
 Pt$d_{z^{2}}$ and  C-$p_{z}$ orbitals become populated to form
 Pt-SWNT bond. This result suggests that in a reverse situation the weakening of the
 Pt-SWNT bond would lead to the transfer of charge from Pt-C bond
 to Pt-H bonds resulting in increased population of
 $d_{xy}$-orbitals in favor of the dissociation.
  Interestingly, exactly the same
configuration has been obtained even  when two H$_{2}$ molecule
approach concomitantly the bare Pt adsorbed on SWNT. Another
configuration related with  two molecularly adsorbed H$_{2}$ is
shown in Fig.~\ref{fig:pt1h2}(b), which appears to be a local
minima on the Born-Oppenheimer surface and $\sim$0.4 eV less
energetics than that in Fig.~\ref{fig:pt1h2}(a). The configuration
of Pt+2H$_{2}$ has remained even after SWNT is removed.

In Fig.~\ref{fig:pt1h2}(c), two H$_{2}$ approaching from both
sites of PtH$_{2}$ already adsorbed on SWNT have been attached by
weak physisorption bonds resulting PtH$_{2}$+2H$_{2}$. Their
distances to Pt atom are relatively larger (2.1\AA{} and 3.2\AA{})
than that occurred for molecularly chemisorption of H$_{2}$. The
latter Pt-H$_{2}$ distance is too long and the binding energy is
$\sim 20$ meV; the binding energy can increase slightly by the VdW
interaction but  the adsorbed molecule can desorb and escape from
Pt at high-temperature. Note that due to weak interaction between
PtH$_{2}$ and both H$_{2}$ molecules, the Pt-C bond becomes
stronger, and consequently H-H distance of PtH$_{2}$ is closed to
be associated in H$_{2}$. The attempts to attach more than three
molecules to the Pt atom have failed. For example, as shown in
Fig.~\ref{fig:pt1h2}(d), from four H$_{2}$ brought at the close
proximity of Pt atom, only three were attached (one dissociatively
chemisorbed, one molecularly chemisorbed, one physisorbed  and the
forth escaped). At the same time the Pt-SWNT bond has weakened and
hence $d_{Pt-C}$ distance has increased to 2.3\AA. We define the
interaction energy between Pt+nH$_{2}$, where n=2 (one of H$_{2}$
is dissociated), 3 and 4 (one of H$_{2}$ is dissociated) and SWNT
in Fig.~\ref{fig:pt1h2}(a-c) as
E[Pt-nH$_{2}$]+E$_{T}$[SWNT]-E$_{T}$[Pt+nH$_{2}$+SWNT]. Here the
total energies are calculated using the same atomic structures in
Fig.~\ref{fig:pt1h2}(a-c). Calculated interaction energies for
each case are 0.68 eV, 1.88 eV and 0.78 eV, respectively.  Using
the similar procedure we also calculated the interaction energy
between Pt+H$_{2}$ (where H$_{2}$ is molecularly adsorbed) and
SWNT in Fig.~\ref{fig:pth}(a) to be 1.93 eV. Note that the
variation of these energies with structure and Pt-C distances
confirm the above arguments related with the dissociation of one
of H$_{2}$ followed by the weakening of the bond between Pt and
SWNT.

\subsection{Adsorption of H2 to a small Pt  cluster  on SWNT}
 As shown in Fig.~\ref{fig:ptall}(a), the situation is different in the case
  of interaction between H$_{2}$ and  a small Pt cluster
 (consisting of a few Pt atoms adsorbed at close proximity). As
H$_{2}$ approaches two Pt atoms on SWNT it starts to dissociate at
a distance $\sim 3.9$\AA{} from the surface of SWNT. The optimized
configuration is shown in Fig.~\ref{fig:ptall}(b) where H-H
molecular bond is broken and  each adsorbed Pt atom  formed a Pt-H
bonds with $d_{Pt-H}$=1.56\AA. Upon chemisorption $d_{Pt-C}$
increased from 2.15\AA{}  to 2.36\AA. The dissociation process
schematically shown in Fig.~\ref{fig:ptall}(c) by plotting the
variation of total energy E$_{T}$ as a function of $z$ for two
different cases. The curve by small circles corresponds to the
total energy of SWNT+Pt$_{2}$ and H$_{2}$ calculated for different
H$_{2}$-tube distance $z$ by keeping the atomic configuration at
$z \rightarrow \infty$ frozen for all $z$. The continuous  curve
is obtained by relaxing the atomic configuration at certain values
of $z$. We see that for $z<4.2$\AA{} E$_{T}$ starts to lower upon
the onset of dissociation. We note very small barrier at about $z
\sim 4.5$\AA. Upon overcoming this energy barrier, the process is
exothermic with an energy gain of $\sim 1.2$ eV. As described in
Fig.~\ref{fig:ptall}(d-f), the adsorption of single H$_{2}$ on a
Pt cluster consisting of three Pt atoms also  results in
dissociation of the molecule. As the size of cluster increased by
inclusion of the third Pt atom, the small potential barrier at
$z\sim 4.5$\AA{} is further lowered, the binding energy increased
to 1.5 eV. Also one of the Pt atoms which binds both H atom is
detaches from the SWNT surface. This situation confirms that
Pt-SWNT bonds are weakened upon the (molecular or dissociative)
adsorption of H$_{2}$ to Pt.

The interaction between Pd atoms adsorbed on SWNT and H$_{2}$
molecule is somehow similar to that with Pt atom. However, the
latter case leads relatively less strong interaction and smaller
binding energies. For example, the interaction between H$_{2}$ and
a single Pd atom adsorbed on SWNT results in a binding between
chemisorption and physisorption with a binding energy of 0.6 eV.
In this case, while the H-H bond length is stretched a little from
the normal value 0.7 \AA~ to 0.8 \AA, the C-Pd bond is stretched
from 2.1 \AA~ to 2.2 \AA. Small changes after the adsorption of
H$_{2}$ are manifestations of relatively weak H$_{2}$-Pd
interaction. In contrast to  adsorbed two Pt atom in
Fig.~\ref{fig:ptall}(a)  breaking the H$_{2}$ molecule, two
adsorbed Pd atoms give rise to chemisorption of molecule with more
stretched H-H bonds.

\section{Conclusions}

In this work we presented a detailed analysis of the interaction
between hydrogen molecule and a SWNT. We found that the binding
energy between H$_{2}$ and outer surface of a bare SWNT is very
weak and the physisorption bond can easily be broken. We showed
that the binding of H$_{2}$ to the outer surface cannot be
enhanced by applying radial deformation to increase curvature
effects at the site facing H$_{2}$ molecule. In contrast, the
interaction between the inner surface of (8,0)  tube and H$_{2}$
is repulsive which can  prevent molecules from entering inside the
tube. The repulsive interaction may turn to be attractive  for
large tube radius.  To promote H$_{2}$ uptake on SWNT surface we
considered functionalized tubes through adsorption of  foreign
atoms. The binding energy of H$_{2}$ on SWNT surface did not
increase by the coadsorption of Li. However, the binding energy
increased if H$_{2}$ is directly attached to adsorbed Li; yet the
nature of the bonding remained physisorption.

The situation with Pt atom, which can make strong chemisorption
bonds with the outer surface of SWNT is found to be interesting
from the point of view of H$_{2}$ storage. We showed that H$_{2}$
molecule can form chemisorption bonds with free Pt as well as Pt
adsorbed on SWNT. Single H$_{2}$ adsorbed on a free Pt atom
dissociates and forms two strong Pt-H bonds. On the other hand,
while single H$_{2}$ molecule  is molecularly chemisorbed to a
single Pt atom at the A-site of SWNT surface,  it can dissociate
if Pt atom adsorbed near the hollow site. Even  the molecular
adsorption of single H$_{2}$ can turn dissociative if a second
H$_{2}$ is molecularly adsorbed to the same Pt atom. The
dissociative adsorption is mediated by the weakening of Pt-C bonds
either due to a specific location of Pt on SWNT or due to the
second H$_{2}$ molecularly adsorbed to Pt. Dissociative adsorption
of single H$_{2}$ to a single Pt atom on the graphite surface
suggests that the dissociation of H$_{2}$ is favored on SWNTs
having large radius. Our analysis suggests that single Pt adsorbed
on SWNT can bind up to two H$_{2}$ molecules with significant
binding energy in the chemisorption range. Beyond two adsorbed
H$_{2}$, additional molecules form weak physisorption bonds with
Pt. Single  Pd atom adsorbed on SWNT exhibits similar effects but
 in relatively weaker manner as compared to  that of Pt. Interesting interaction
between H$_{2}$ and Pt and resulting bonding mechanisms justifies
similar investigations of SWNTs functionalized by other transition
elements (in particular Ti, Ni, Cr, V) for a higher $wp$ H$_{2}$
storage.

\begin{acknowledgments}
 SC acknowledges partial support from
Academy of Science of Turkey.
\end{acknowledgments}


\begin{thebibliography}{99}
\bibitem{hydrogen} Since, the bond distance of H$_{2}$ is extended
by 20$\%$, the H-H interaction is weaker than that in free
H$_{2}$. For d$_{H-H}\sim$1\AA{} the adsorption may be considered
as an intermediate stage between molecular and atomic adsorption.
\end{thebibliography}
\end{document}